\documentclass[prd,amsmath,amssymb,showpacs,superscriptaddress,nofootinbib,twocolumn]{revtex4-1}
\usepackage{graphicx}
\usepackage{epsfig,graphics,subfigure,psfrag,amsmath,amssymb}
\usepackage{lineno}
\usepackage{dcolumn}
\usepackage{bm}

\usepackage{overpic}
\usepackage[english]{babel}
\usepackage{color}
\usepackage{multirow}
\usepackage{colortbl}
\usepackage{epstopdf}
\usepackage[colorlinks,linkcolor=black,anchorcolor=blue,citecolor=blue]{hyperref}
\usepackage{gensymb}
\usepackage{ulem}

\usepackage[
total={6.5in,8.75in}, top=1.2in, left=0.9in, 
]{geometry}
\begin{document}

\title{\texorpdfstring{\boldmath
First observation of the decay $\chi_{cJ}\to \Sigma^{+}\bar{p}K_{S}^{0}+c.c. ~(J = 0, 1, 2)$ } {}}


\author{M.~Ablikim$^{1}$, M.~N.~Achasov$^{10,d}$, P.~Adlarson$^{60}$, S. ~Ahmed$^{15}$, M.~Albrecht$^{4}$, M.~Alekseev$^{59A,59C}$, A.~Amoroso$^{59A,59C}$, F.~F.~An$^{1}$, Q.~An$^{56,44}$, Y.~Bai$^{43}$
% ...
\\
 \vspace{0.2cm}
 (BESIII Collaboration)\\
 \vspace{0.2cm} {\it
$^{1}$ Institute of High Energy Physics, Beijing 100049, People's Republic of China\\
% ...
$^{4}$ Bochum Ruhr-University, D-44780 Bochum, Germany\\
% ...
$^{10}$ G.I. Budker Institute of Nuclear Physics SB RAS (BINP), Novosibirsk 630090, Russia\\
% ...
$^{15}$ Helmholtz Institute Mainz, Johann-Joachim-Becher-Weg 45, D-55099 Mainz, Germany\\
% ...
$^{43}$ Southeast University, Nanjing 211100, People's Republic of China\\
$^{44}$ State Key Laboratory of Particle Detection and Electronics, Beijing 100049, Hefei 230026, People's Republic of China\\
% ...
$^{56}$ University of Science and Technology of China, Hefei 230026, People's Republic of China\\
% ...
$^{59}$ (A)University of Turin, I-10125, Turin, Italy; (B)University of Eastern Piedmont, I-15121, Alessandria, Italy; (C)INFN, I-10125, Turin, Italy\\
$^{60}$ Uppsala University, Box 516, SE-75120 Uppsala, Sweden\\
% ...
$^{d}$ Also at the Novosibirsk State University, Novosibirsk, 630090, Russia\\
% ...
}
}


\begin{abstract}
  Using E1 radiative transitions $\psi(3686) \to \gamma\chi_{cJ}$
  from a sample of $(448.1 \pm 2.9)\times10^{6}$ $\psi(3686)$
  events collected with the BESIII detector, the decays $\chi_{cJ}\to
  \Sigma^{+}\bar{p}K_{S}^{0}+c.c.~(J = 0, 1, 2)$ are studied.  The
  decay branching fractions  are measured to be
  $\mathcal{B}(\chi_{c0}\to \Sigma^{+}\bar{p}K_{S}^{0}+c.c.) = (3.52
  \pm 0.19\pm 0.21)\times10^{-4}$, $\mathcal{B}(\chi_{c1}\to
  \Sigma^{+}\bar{p}K_{S}^{0}+c.c.) = (1.53 \pm 0.10\pm
  0.08)\times10^{-4}$, and $\mathcal{B}(\chi_{c2}\to
  \Sigma^{+}\bar{p}K_{S}^{0}+c.c.) = (8.25 \pm 0.83 \pm
  0.49)\times10^{-5}$, where the first and second uncertainties are
  the statistical and systematic ones, respectively.  No evident
  intermediate resonances are observed in the studied processes.
\end{abstract}

\maketitle

\section{Introduction}
The first charmonium states with $J^{PC}= J^{++}$ discovered after the
$J/\psi$ and $\psi (3686)$ were the $\chi_{cJ}~(J = 0, 1, 2)$
particles.  Quarkonium systems,
especially charm anti-charm states, are regarded as a unique
laboratory to study the interplay between perturbative and
nonperturbative effects in quantum chromodynamics (QCD). Experimental
studies of charmonium decays can test QCD and QCD-based effective
field theory calculations.  The $\chi_{cJ}$ states belong to the
charmonium $P$-wave spin triplet, and therefore cannot be produced
via a single virtual-photon exchange in electron-positron annihilations
as are the $J/\psi$ and
$\psi (3686)$.  Until now the understanding of these states has been
limited by the availability of experimental data. The world's largest
data set of $\psi(3686)$ events ~\cite{psi'data} collected with the
BESIII~\cite{BES3} detector, provides a unique opportunity for
detailed studies of $\chi_{cJ}$ decays, since they are copiously
produced in $\psi (3686)$ radiative transitions with branching
fractions of about 9$\%$ each~\cite{PDG2016}.

Many excited baryon states have been discovered by BaBar, Belle, CLEO,
BESIII, and other experiments in the past decades~\cite{PDG2016}, but
the overall picture of these states is still unclear. While many
predicted states have not yet been observed, many states that do not
agree with quark model predictions are observed (for a review see Ref.~\cite{review}). Therefore the search
for new excited baryon states is important to improve knowledge of
the baryon spectrum and the understanding of the underlying processes
which describe confinement in the nonperturbative QCD
regime. Experimentally, exclusive decays of $\chi_{cJ}$ to baryon
anti-baryon ($B\bar{B}$) pairs, such as $p\bar{p}$,
$\Sigma\bar{\Sigma}$, $\Lambda \bar{\Lambda}$~\cite{ppbar,lambda,Sigma,BB}, have been
investigated. However, there are only a few experimental studies of
$\chi_{cJ}$ to $B\bar{B}M$ ($M$ stands for meson). These channels are
ideal to search for new excited baryons in intermediate
states, which decay into $\bar{B}M$ and $BM$.

This paper reports the first measurements of the branching fractions
of $\chi_{cJ}\to\Sigma^{+}\bar{p}K_{S}^{0}+c.c.$ via the radiative
transition $\psi(3686) \to \gamma\chi_{cJ}$, where $\Sigma^{+}\to p
\pi^{0}, K_{S}^{0}\to \pi^{+}\pi^{-}$, and $\pi^{0} \to
\gamma\gamma$. The charge-conjugate state ($c.c.$) is included unless
otherwise stated.  We also report on a search for possible excited
baryon states in the invariant-mass spectra of $\bar{p}K_{S}^{0}$,
and $\Sigma^{+}K_{S}^{0}$.

\section{\texorpdfstring{BESIII Detector}{Detector and MC Simulation}}
The BESIII detector is a magnetic spectrometer located at the Beijing
Electron Positron Collider (BEPCII)~\cite{bepc}. The cylindrical core
of the BESIII detector consists of a helium-based multilayer drift
chamber (MDC), a plastic scintillator time-of-flight system (TOF), and
a CsI (Tl) electromagnetic calorimeter (EMC), which are all enclosed
in a superconducting solenoidal magnet providing a 1.0~T magnetic
field.  The solenoid is supported by an octagonal flux-return yoke
with resistive plate counter muon identifier modules interleaved with
steel. The acceptance of charged particles and photons is 93\% over
4$\pi$ solid angle. The charged-particle momentum resolution at 1~GeV
is 0.5\%, and the $dE/dx$ resolution is 6\% for the electrons from
Bhabha scattering at 1~GeV. The EMC measures photon energies with a
resolution of 2.5\% (5\%) at 1~GeV in the barrel (end-cap) region. The
time resolution of the TOF barrel part is 68~ps, while that of the
end-cap part is 110 ps.

\section{\texorpdfstring{Data Set and Monte Carlo Simulation}{}}
This analysis is based on a sample of $(448.1\pm2.9)\times10^{6}$
$\psi(3686)$ events~\cite{psi'data} collected with the BESIII
detector.

{\sc{geant4}}-based~\cite{GEANT} Monte Carlo (MC) simulation data are
used to determine detector efficiency, optimize event selection, and
estimate background contributions.  Inclusive MC samples were produced
to determine contributions from dominant background channels.  The
production of the initial $\psi(3686)$ resonance is simulated by the
MC event generator {\sc{kkmc}}~\cite{kkmc,kkmc2}, and the known decay
modes are modeled with {\sc{evtgen}}~\cite{evtgen,evtgen2} using the
branching fractions summarized and averaged by the Particle Data Group
(PDG)~\cite{PDG2016}, while the remaining unknown decays are generated
by {\sc{lundcharm}}~\cite{lundcharm}. The final states are propagated
through the detector system using {\sc{geant4}} software.

In addition, for the optimization of the selection criteria and the
determination of the efficiency, exclusive MC data sets with
$4\times10^5$ events are generated for each signal mode. Here, the
$\psi(3686) \to \gamma\chi_{cJ}$ decay is generated assuming an E1
transition~\cite{signalMC1,signalMC2}, where the photon polar angle
$\theta$ in the $e^+e^-$ center-of-mass frame is distributed according
to ($1+\lambda\cos^2\theta$). For $J$ = 0, 1, and 2, $\lambda$ is set
to $1, -\frac{1}{3}$, and $\frac{1}{13}$, respectively. The decays
$\chi_{cJ}\to\Sigma^{+}\bar{p}K_{S}^{0},\Sigma^{+}\to p \pi^{0},
K_{S}^{0}\to \pi^{+}\pi^{-},\pi^{0}\to \gamma\gamma $ are generated by
using the phase-space model (PHSP).
\section{Data analysis}

For the reaction channel $\psi(3686)\rightarrow \gamma\chi_{cJ}$, with
$\chi_{cJ}\to\Sigma^{+}\bar{p}K_{S}^{0},\Sigma^{+}\to p \pi^{0},
\pi^{0}\to \gamma \gamma$, and $K_{S}^{0}\to \pi^{+}\pi^{-}$, the
final-state particles are
$p\bar{p}\pi^+\pi^-\gamma\gamma\gamma$. Charged tracks must be in the
active region of the MDC, corresponding to $|\cos\theta|<$ 0.93, where
$\theta$ is the polar angle of the charged track with respect to the
beam direction.  For the anti-proton ($\bar{p}$), the point of closest
approach to the interaction point (IP) must be within $\pm$1 cm in the
plane perpendicular to the beam ($R_{xy}$) and $\pm$10 cm along the
beam direction ($V_z$). Due to the long lifetime of the $K_{S}^{0}$
and $\Sigma^{+}$, there is no requirement on $R_{xy}$ or $V_z$ for
the track candidates used to form the $K_{S}^{0}$ or $\Sigma^{+}$
candidates.  Photon candidates are reconstructed by summing the energy
deposition in the EMC crystals produced by the electromagnetic
showers. The minimum energy necessary for counting a photon as a
photon candidate is 25~MeV for barrel showers ($|\cos\theta|<0.8$) and
50~MeV for end-cap showers ($0.86<|\cos\theta|<0.92$). To eliminate
showers originating from charged particles, a photon cluster must be
separated by at least $10^\circ$ from any charged track. The timing of
the shower is required to be within 700~ns from the reconstructed
event start time to suppress noise and energy deposits unrelated to
the event. Events with two positively charged tracks, two negatively
charged tracks, and at least three good photons are selected for
further analysis.  The TOF (both end-cap and barrel) and $dE/dx$
measurements for each charged track are used to calculate the p-value
based on the $\chi_{\rm PID}^2$ values for the hypotheses that a track
is a pion, kaon, or proton.  Two oppositely charged tracks are
identified as a proton/anti-proton pair if their proton hypothesis
p-values are greater than their kaon or pion hypothesis p-values.  The
remaining charged tracks are considered as pions by default. The
numbers of protons and anti-protons as well as the negatively and
positively charged pions should be equal to one.

The $K_{S}^{0}$ candidate is reconstructed with a pair of oppositely
charged pions.  To suppress events from combinatorial background
contributions, we require that the $\pi^{+}\pi^{-}$ pair is produced
at a common vertex~\cite{senvt}.

\begin{figure}[htbp]
    \centering
    \vskip -0.0cm
    \hskip -0.0cm \mbox{
    \begin{overpic}[width=7.5cm,height=5.0cm,angle=0]{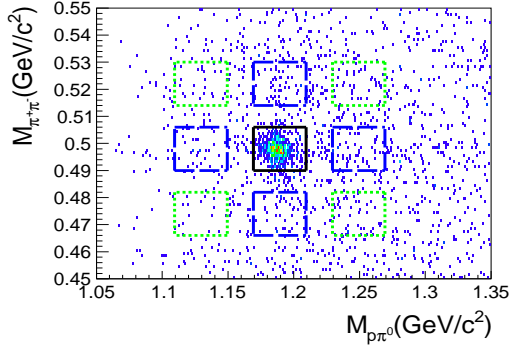}
    \end{overpic}} {\caption{ The distribution of the $\pi^{+}\pi^{-}$
      invariant mass versus the $p\pi^{0}$ invariant mass.  The  black solid box in the center is the signal region, the blue long
      dashed boxes show the $K_{S}^{0}$ and $\Sigma^{+}$ mass sideband
      regions, and the green dashed boxes are the events from
      non-$K_{S}^{0}$ and non-$\Sigma^{+}$ candidates.  }
    \label{2D}}
    \end{figure}

Next a four-constraint (4C) kinematic fit imposing energy-momentum
conservation is performed under the
$p\bar{p}\pi^+\pi^-\gamma\gamma\gamma$ hypothesis. If there are more
than three photon candidates in an event, the combination with the
smallest $\chi^{2}_{\rm 4C}$ is retained, and its $\chi^{2}_{\rm 4C}$
is required to be less than those for the
$p\bar{p}\pi^{+}\pi^{-}\gamma\gamma$ and
$p\bar{p}\pi^{+}\pi^{-}\gamma\gamma\gamma\gamma$ hypotheses.  The
value of $\chi^{2}_{\rm 4C}$ is required to be less than 50.  For the
selected signal candidates, the $\gamma\gamma$ combination
($\gamma_1\gamma_2$) with an invariant mass closest to the $\pi^{0}$ mass
is reconstructed as $\pi^{0}$ candidate, and the remaining one
($\gamma_3$) is considered to be the radiative photon from the
$\psi(3686)$ decay. The $\gamma\gamma$ invariant mass is required to
satisfy $|M_{\gamma\gamma}-m_{\pi^{0}}|< 15 $ MeV/$c^{2}$. Here and
throughout the text, $M_{i}$ represents a measured invariant mass and
$m_{i}$ represents the nominal mass of the particle(s)
$i$~\cite{PDG2016}. To reduce background events with
$\bar{\Lambda}\rightarrow\bar{p}\pi^+$ ,
$|M_{\bar{p}\pi^+}-m_{\Lambda}|>6 $ MeV/$c^{2}$ is required.
Figure ~\ref{2D} shows the scatter plot of the $\pi^{+}\pi^{-}$
invariant mass versus the $p\pi^{0}$ invariant mass of data.  To
select events which contain both a $K_{S}^{0}$ and a $\Sigma^+$
candidate, $|M_{\pi^{+}\pi^{-}}-m_{K_{S}^{0}}|<8$ MeV/$c^{2}$ and
$|M_{p\pi^{0}}-m_{\Sigma^+}|<20$ MeV/$c^{2}$ are required ( black solid box in Fig.~\ref{2D}). The widths of the mass intervals are
chosen to be 3 times the invariant-mass resolution.

 \begin{figure}[htbp]
    \centering
    \vskip -0.0cm
    \hskip -0.0cm \mbox{
    \begin{overpic}[width=7.5cm,height=5.0cm,angle=0]{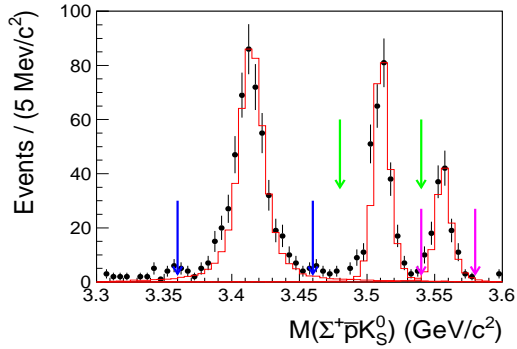}
    \end{overpic}} {\caption{The $\Sigma^{+}\bar{p} K_{S}^{0}$
      invariant-mass distribution in the vicinity of the $\chi_{cJ}$
      states. Dots with error bars are data, the red solid line
      histogram is the $\chi_{cJ}$ line shape from the MC simulation,
      and the arrows indicate the $\chi_{c0}$, $\chi_{c1}$, and
      $\chi_{c2}$ signal regions.  }
    \label{chicJm}}
    \end{figure}

    The $\Sigma^{+}\bar{p}K_{S}^{0}$ invariant-mass distributions of
    the 937 events that passed all selection criteria and the MC
    simulated events are shown in Fig.~\ref{chicJm}. Clear
    signals are observed in the $\chi_{c0}$, $\chi_{c1}$, and
    $\chi_{c2}$ mass regions. The $\chi_{c0}$, $\chi_{c1}$,
    and $\chi_{c2}$ decays are defined as [3.36, 3.46], [3.48,
    3.54], and [3.54, 3.58] GeV/$c^{2}$, respectively, as
    indicated with arrows in Fig.~\ref{chicJm}.

\begin{figure}[htbp]
    \centering
    \vskip -0.0cm
    \hskip -10.0cm

    \subfigure{
\includegraphics[width=2.5in,height=1.4in]{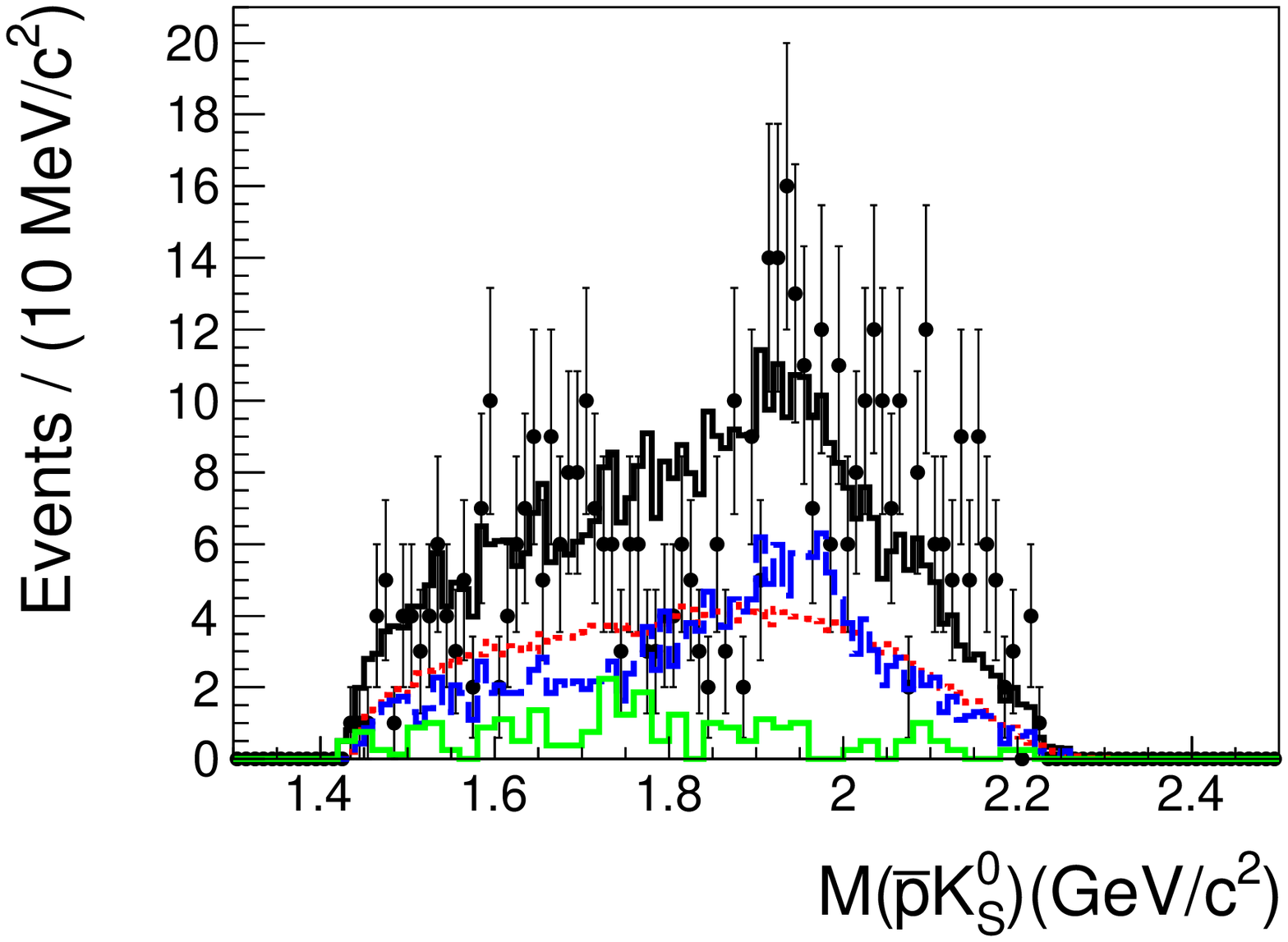}\put(-140,80){(a)}
}

\vspace{-0.5cm}
\subfigure{
\includegraphics[width=2.5in,height=1.4in]{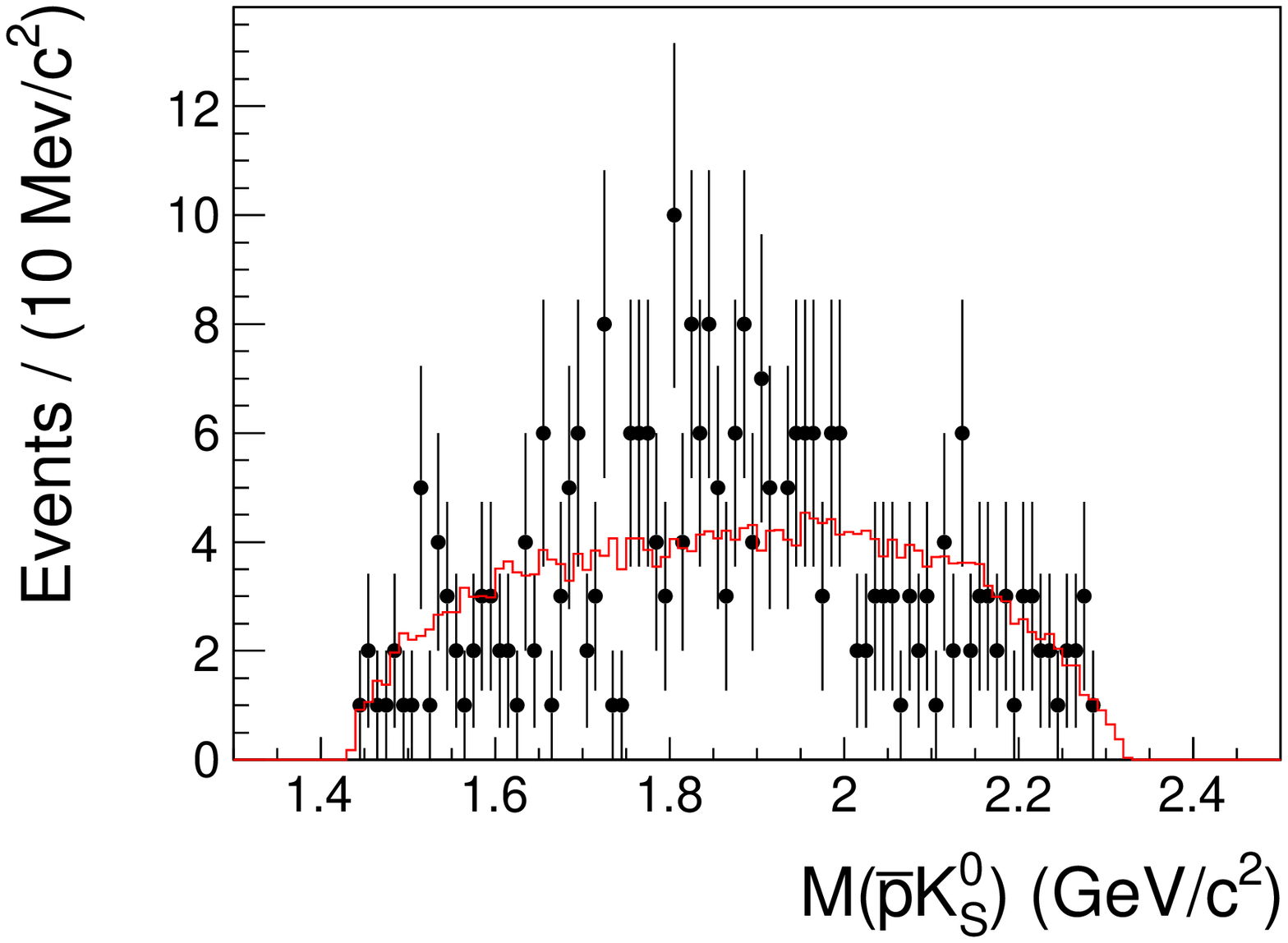}\put(-140,80){(b)}
}

\vspace{-0.5cm}
\subfigure{
\includegraphics[width=2.5in,height=1.4in]{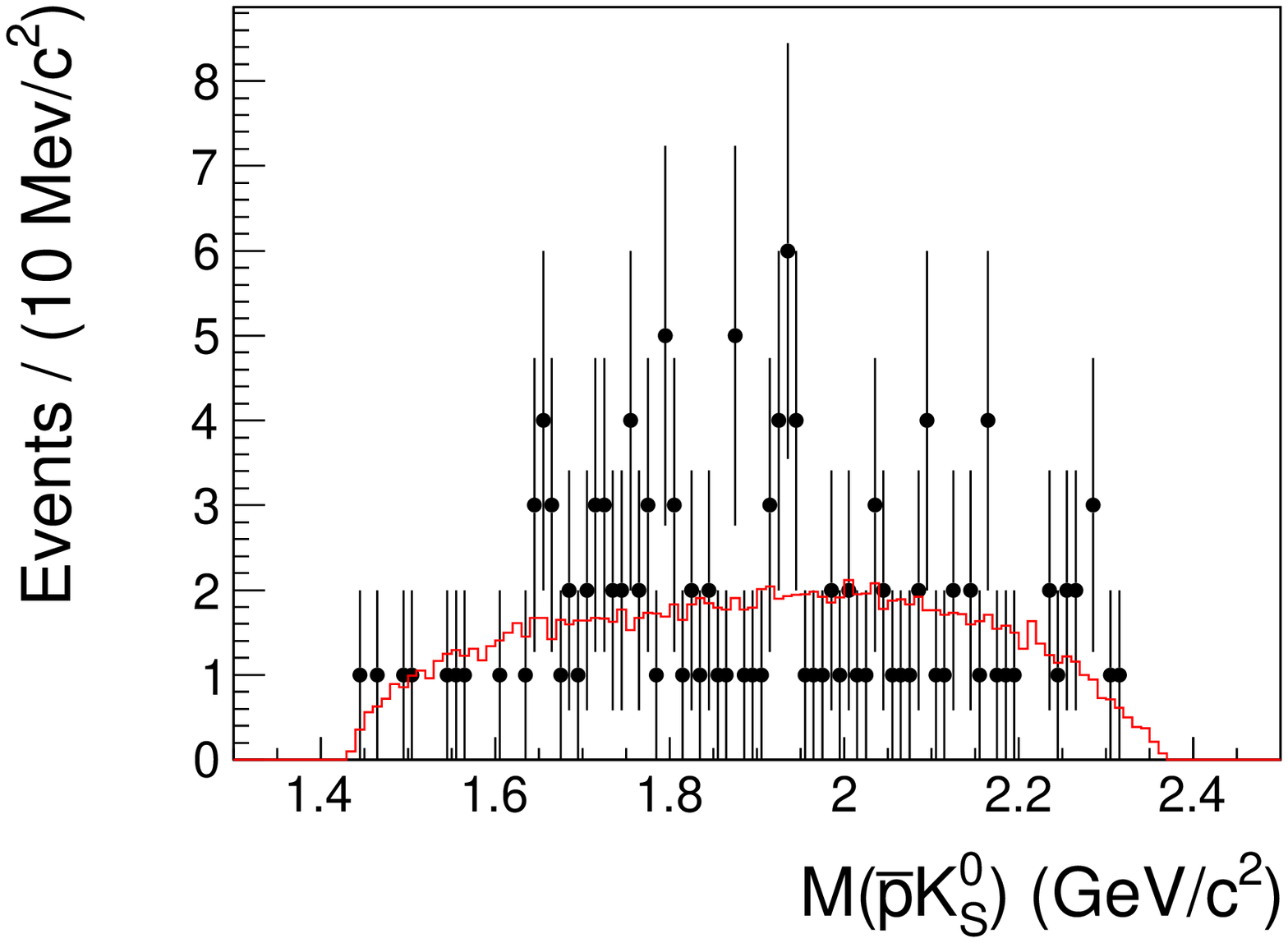}\put(-140,80){(c)}
}

{\caption{ The $\bar{p}K^{0}_{S}$ invariant-mass distributions in the
    (a) $\chi_{c0}$, (b) $\chi_{c1}$, and (c) $\chi_{c2}$ signal
    regions. The dots with error bars are data, the red lines are the
    contributions from the corresponding MC simulations based on the
    phase-space model. For plot (a), the black solid line is the fit
    result, the blue long-dashed curve is the contribution from $
    \chi_{c0}\to \Sigma^{+}\bar\Sigma(1940)^{-}$, and the green solid
    line is the contribution from the normalized $K_{S}^{0}$ and
    $\Sigma^{+}$ mass sideband regions.  }
    \label{PKS}}
\end{figure}

    A hint of a structure in the invariant-mass distribution of the
    $\bar{p}K_{S}^{0}$ subsystem in the $\chi_{c0}$ signal region can
    be seen in Fig.~\ref{PKS}(a). Considering the width and mass, it
    is most likely the $\bar\Sigma(1940)^{-}$ with $M = 1940$
    MeV/$c^2$, $\Gamma$ = 220 MeV, and $I(J^{P}) = 1
    (\frac{3}{2}^-)$~\cite{PDG2016}.  Other excited $\Sigma^*$ states
    are most likely excluded because their widths are much larger.
    For the fit to the invariant-mass distribution
    $M_{\bar{p}K_{S}^{0}}$, several contributions are considered, namely the
    line shape from the phase-space model, the normalized $K_{S}^{0}$
    and $\Sigma^{+}$ mass sidebands in the $\chi_{c0}$ signal region
    (described in detail in the background analysis), and the
    $\bar\Sigma(1940)^{-}$ signal from the MC simulation, where the
    mass and width of $\bar\Sigma(1940)^{-}$ are fixed to the world
    average values~\cite{PDG2016}.  To estimate the statistical signal
    significance of the $\bar\Sigma(1940)^{-}$ contribution, we use
    the quantity $\sqrt{-2\ln(\mathcal{L}_0/\mathcal{L}_{\rm max})}$,
    where $\mathcal{L}_0$ and $\mathcal{L}_{\rm max}$ are the
    likelihoods of the fits without and with $\bar\Sigma(1940)^{-}$
    signal, respectively. The statistical significance of the
    $\bar\Sigma(1940)^{-}$ signal is obtained to be $3.2\sigma$. The
    signal significance is reduced to $2.3\sigma$ if the width of
    $\bar\Sigma(1940)^{-}$ is taken as the lower value of 150
    MeV~\cite{PDG2016}.  For all other invariant-mass distributions of
    the two-body subsystems, the description using the phase-space
    model is in good agreement with data. For example, the
    $\bar{p}K^{0}_{S}$ mass distributions from data and MC simulations
    in the $\chi_{c1}$ and $\chi_{c2}$ signal regions are shown in
    Figs.~\ref{PKS}(b) and~\ref{PKS}(c).

\begin{figure}[htbp]
    \centering
    \vskip 0.0cm
    \hskip 0.0cm
    \begin{overpic}[width=7.5cm,height=5.0cm,angle=0]{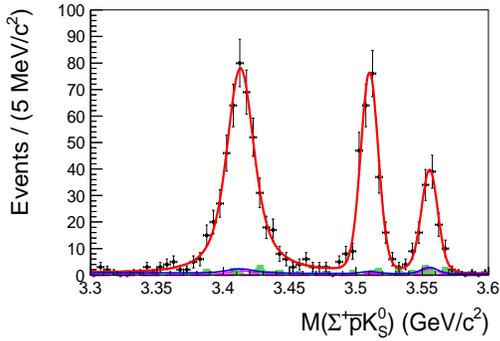}
    \end{overpic}
    {\caption{ Fit to the $\Sigma^+\bar{p}K_{S}^{0}$ invariant-mass
        distribution in the $\chi_{cJ}$ mass region of [3.3, 3.6]
        GeV/$c^{2}$. Dots with error bars are data, the red solid
        curve shows the result of the unbinned maximum-likelihood fit,
        the green-shaded histograms are the events from the normalized
        $K_{S}^{0}$ and $\Sigma^{+}$ mass sidebands, the blue solid
        line is sum of the peaking and flat background components, and
        the violet long dashed curve is the contribution of the
        peaking background normalized according to the sideband
        events.  }
    \label{fitchicJ}}
\end{figure}

Possible background contributions are studied with the inclusive MC
sample of $5.06\times 10^8$ simulated $\psi(3686)$ decays. Peaking
background contributions in the $\chi_{cJ}$ mass regions are dominated
by the channels $\chi_{cJ} \rightarrow
\bar{\Delta}^{-}\pi^{+}\Delta^0 (\bar{\Delta}^{-} \rightarrow
\bar{p}\pi^{0}, \Delta^0 \rightarrow p \pi^{-})$ and $\chi_{cJ}
\rightarrow p \bar{p} \rho^{+} \pi^{-} (\rho^{+} \rightarrow \pi^{+}
\pi^{0})$.  Other background events, mainly from the channels
$\psi(3686) \rightarrow \Sigma^+ \bar{p} K^{*}(\Sigma^+ \rightarrow p
\pi^{0} , K^{*} \rightarrow K_{S}^{0} \pi^{0} , K^{0}_{S} \rightarrow
\pi^{+} \pi^{-})$, $\psi(3686) \rightarrow K_{S}^{0} \bar{\Delta}^-
\Sigma^+(\bar{\Delta}^- \rightarrow \bar{p} \pi^{0} , \Sigma^+
\rightarrow p \pi^{0} , K^{0}_{S} \rightarrow \pi^{+} \pi^{-}) $ and
$\psi(3686) \rightarrow J/\psi \pi^{0} \pi^{0}(J/\psi \rightarrow p
\bar{\Delta}^0 \pi^{-} , \bar{\Delta}^0 \rightarrow \bar{p} \pi^{+} )
$ are not peaking in the $\chi_{cJ}$ mass regions. The amount of
background events is estimated by using the normalized $K_{S}^{0}$ and
$\Sigma^{+}$ mass sideband events, as shown in Fig.~\ref{2D}. The blue
long dashed boxes are the selected $K_{S}^{0}$ mass sidebands
($1.1694<M_{p\pi^{0}}<1.2094$ GeV/$c^{2}$, $0.466
<M_{\pi^{+}\pi^{-}}<0.482$ GeV/$c^{2}$ and
$0.514<M_{\pi^{+}\pi^{-}}<0.530$ GeV/$c^{2}$) and the $\Sigma^{+}$
mass sidebands ($0.49 <M_{K_{S}^{0}}<0.506$ GeV/$c^{2}$,
$1.1094<M_{p\pi^{0}}<1.1494$ GeV/$c^{2}$ and
$1.2294<M_{p\pi^{0}}<1.2694$ GeV/$c^{2}$), and the green dashed boxes
are those from non-$K_{S}^{0}$ and non-$\Sigma^{+}$ sidebands
($1.1094<M_{p\pi^{0}}<1.1494$ GeV/$c^{2}$ and
$1.2294<M_{p\pi^{0}}<1.2694$ GeV/$c^{2}$, $0.466
<M_{\pi^{+}\pi^{-}}<0.482$ GeV/$c^{2}$ and
$0.514<M_{\pi^{+}\pi^{-}}<0.530$ GeV/$c^{2}$). The normalized
background contribution in the $\chi_{cJ}$ mass regions is estimated
as half of the total number of events in the four blue sideband
regions minus one quarter of the total number of events in the four
green sideband regions of Fig.~\ref{2D}, and shown as a green-shaded
histogram in Fig.~\ref{fitchicJ}.

An unbinned maximum-likelihood fit to the $\Sigma^+\bar{p}K_{S}^{0}$
invariant-mass distribution is performed for the total selected signal
candidates, as shown in Fig.~\ref{fitchicJ}. The complete formula for
the fit is $PDF_{\rm total} = N_1 \times PDF_{\rm signal}+ N_2 \times
PDF_{\rm peaking bkg}+ N_3 \times PDF_{\rm flat bkg}$. The parameters
$N_1$ and $N_3$ are free, and $N_2$ is fixed to the number of the
events determined from the $K_{S}^{0}$ and $\Sigma^{+}$ mass sidebands.

Here, $PDF_{\rm signal}$ is the sum of the signal line shapes of the
three $\chi_{cJ}$ resonances each convolved with a Gaussian function
related to the $\chi_{cJ}$ mass resolution, where the width of the Guassian function is fixed to each of the MC simulated value. The line shape of each
resonance is described by:
\begin{equation}
  \begin{aligned}
    PDF_{\rm signal, \chi_{cJ}} = BW(M)\times E_{\gamma}^{3}\times
    D(E_{\gamma}),
     \end{aligned}
\end{equation}
where $M$ is the $\Sigma^+\bar{p}K_{S}^{0}$ invariant mass,
$BW(M)=\frac{1}{(M-m_{\chi_{cJ}})^2+0.25\Gamma_{\chi_{cJ}}^2}$ is the
Breit-Wigner function, with $m_{\chi_{cJ}}$ and $\Gamma_{\chi_{cJ}}$
the mass and width of the corresponding $\chi _{cJ}$,
$E_{\gamma}=\frac{m_{\psi(3686)}^{2}-M^2}{2m_{\psi(3686)}}$ is the
energy of the transition photon in the rest frame of $\psi(3686)$ and
$D(E_{\gamma})$ is the damping factor which suppresses the divergent
tail due to the $E_{\gamma}^{3}$ dependence of $PDF_{\rm signal}$. It
is described by $\exp(-E_{\gamma}^{2}/8\beta^{2})$ where $\beta$ is
one of the free parameters in the fit. For all three resonances the
same $\beta$ value is required. The fit result $\beta=(68.7 \pm 13.0)$
MeV is consistent with the value measured by the CLEO
experiment~\cite{CLEO}.

The peaking background component $PDF_{\rm peaking bkg}$ is the same
as the signal distribution.  It is used to describe the distribution
of the normalized events from the $K_{S}^{0}$ and $\Sigma^{+}$ mass
sidebands where clearly the three $\chi_{cJ}$ resonances can be
identified.  The $PDF_{\rm flat bkg}$ is described by a first-order
polynomial.

For the unbinned maximum-likelihood fit, $\beta$, the masses and
widths of the $\chi_{cJ}$ resonances, and the two coefficients of the
polynomial are taken as free parameters. The event yields of
the fitted $\chi_{cJ}\to \Sigma^{+}\bar{p}K_{S}^{0}$ signals are
listed in Table~\ref{result}.

\begin {table*}[htp]
  \centering {\caption {Number of signal events ($N^{\chi_{cJ}}_{\rm
        obs}$), detection efficiency ($\epsilon$), and branching
      fractions $\mathcal{B}(\chi_{cJ}\to \Sigma^{+}\bar{p}K_{S}^{0}
      )$, where the first uncertainty is statistical and the second is
      systematic. 
    }
    \label{result}}
    \begin{tabular}{l|c|c|c} \hline \hline
       Mode &  $N^{\chi_{cJ}}_{\rm obs}$ & $\epsilon(\%)$&$\mathcal{B}(\chi_{cJ}\to \Sigma^{+}\bar{p}K_{S}^{0}  )$\\ \hline
      $\chi_{c0}\to \Sigma^{+}\bar{p}K_{S}^{0}$  &$493 \pm 26$ &$9.05 \pm 0.05$&$ (3.52 \pm 0.19\pm 0.21)\times10^{-4}$\\ \hline
      $\chi_{c1}\to \Sigma^{+}\bar{p}K_{S}^{0}$ & $258 \pm 17$ &$10.96 \pm 0.05$& $(1.53\pm 0.10\pm 0.08)\times10^{-4}$\\ \hline
     $\chi_{c2}\to \Sigma^{+}\bar{p}K_{S}^{0}$ & $129 \pm 13$  &$10.40 \pm 0.05$&$(8.25\pm 0.83 \pm 0.49)\times10^{-5}$\\ \hline\hline
    \end{tabular}
\end{table*}

The branching fractions for $\chi_{cJ}\to \Sigma^{+}\bar{p}K_{S}^{0}$
are calculated by \begin{equation}
      \begin{aligned}
      \mathcal{B}(\chi_{cJ} \to \Sigma^{+}\bar{p}K_{S}^{0}  )
      =\frac{N^{\chi_{cJ}}_{\rm obs}}{N_{\psi(3686)}\times\epsilon\times \prod_{j}\mathcal{B}_{j}},
\end{aligned}
\end{equation}
where $N_{\psi(3686)}$ is the total number of $\psi(3686)$ events,
$\epsilon$ is the corresponding detection efficiency as listed in
Table ~\ref{result}, which is obtained by weighting the simulated
Dalitz plot distribution with the distribution from data, and
$\prod_{j}\mathcal{B}_{j} = \mathcal{B}(\psi(3686)\to
\gamma\chi_{cJ})\times\mathcal{B}(\Sigma^{+}\to p
\pi^{0})\times\mathcal{B}(K_{S}^{0}\to
\pi^{+}\pi^{-})\times\mathcal{B}(\pi^{0}\to\gamma\gamma)$, where the
branching fractions are taken from the PDG~\cite{PDG2016}. The results of
the branching-fraction calculation for the decays $\chi_{cJ}\to
\Sigma^{+}\bar{p}K_{S}^{0}$ are also listed in Table~\ref{result} with
statistical and systematic uncertainties.

\section{Systematic Uncertainties}
The systematic uncertainties on the $\chi_{cJ}\to
\Sigma^{+}\bar{p}K_{S}^{0}$ branching-fraction measurements are listed
in Table~\ref{summary_of_syserr}.

The systematic uncertainty of the photon-detection efficiency is
studied by considering the decay
$J/\psi\to\pi^+\pi^-\pi^0$~\cite{photon} and is about 1\% for each
photon, so 3\% is assigned for the three photons in the final
states.

The uncertainty related to the particle identification (PID) and
tracking of the proton and anti-proton is studied with the control samples
of $J/\psi$ and $\psi(3686)\to p\bar{p}\pi^{+}\pi^{-}$\cite{proton}.
The average differences of efficiencies between MC simulations and
data are 0.4\%, 0.4\%, and 0.3\% for the proton from $\chi_{c0}$,
$\chi_{c1}$, and $\chi_{c2}$ decays, respectively, with the transverse
momentum and angle region of our signal channel considered.  Similarly
for $\bar{p}$, they are 0.4\%, 0.3\%, and 0.3\%, respectively, so the
uncertainties on the proton and anti-proton pair PID and tracking are
0.6$\%$, 0.5\%, and 0.4\% for $\chi_{c0}$, $\chi_{c1}$, and
$\chi_{c2}$ decays, respectively.

The uncertainty associated with the 4C kinematic fit comes from the
inconsistency between data and MC simulation, as described in detail
in Ref.~\cite{refsmear}.  In this analysis, we take the efficiency
with the correction as the nominal value, and the differences between the
efficiencies with and without correction, 0.4\%, 0.4\%, and 0.3\% for
$\chi_{c0}$, $\chi_{c1}$, and $\chi_{c2}$, respectively, as the
systematic uncertainties from the kinematic fit.

The uncertainty associated with the $K_S^{0}$ reconstruction is
studied using $J/\psi \to K^{*}(892)^{\pm}K^{\mp}$,
$K^{*}(892)^{\pm}\to K_S^{0}\pi^{\pm}$ and $J/\psi \to \phi
K_S^{0}K^{\pm}\pi^{\mp}$ control samples and is estimated to be
1.2\%~\cite{ksre}.

The uncertainty related with the $\pi^{0}$ ($K_S^{0}$, $\Sigma^{+}$) mass
window requirement is studied by fitting the $\pi^{0}$ ($K_S^{0}$,
$\Sigma^{+}$) mass distributions of data and signal MC simulation with
a free Crystal Ball (Gaussian, Gaussian) function and a first-order
Chebyshev polynomial function. We obtained the selection efficiency of the
$\pi^0$ ($K_S^{0}$, $\Sigma^{+}$) mass region, which is the ratio of
the numbers of $\pi^0$ ($K_S^{0}$, $\Sigma^{+}$) events with and
without the $\pi^0$ ($K_S^{0}$, $\Sigma^{+}$) mass window, determined
by integrating the fitted signal shape. The difference in efficiency
between data and MC simulation, $ 0.3\%$ ($ 0.3\%$, $ 0.1\%$), is
assigned as the systematic uncertainty.  The systematic uncertainty
from the veto of the $\Lambda$ mass window is negligible due to the
high detection efficiency.

The uncertainty of the detection efficiency is studied
by changing the number of bins in the Dalitz plot. The
maximum differences of the signal detection efficiency, 1.0\%, 0.5\%
and 0.4\% , are taken as uncertainties for $\chi_{c0}$, $\chi_{c1}$,
and $\chi_{c2}$ decays, respectively.
The uncertainty of assuming $\psi(3686)\to\gamma\chi_{c1}(\chi_{c2})$ as
pure E1 transition is studied by considering the contribution from higher order multiple amplitudes~\cite{M2} in the MC simulation, the differences of the efficiency, 0.8\% for $\chi_{c1}$ and 0.2\% for $\chi_{c2}$, are taken as the systematic uncertainties.  For $\chi_{c0}\to
\Sigma^{+}\bar{p}K_{S}^{0}$, there is a possible structure in the
$\bar{p}K_{S}^{0}$ invariant distribution.  The corresponding
systematic uncertainty is estimated by mixing $\chi_{c0}\to
\Sigma^{+}\bar{\Sigma}(1940)^{-}$ MC sample and the PHSP signal MC
sample in a proportion, which is obtained from fitting
$M_{\bar{p}K_{S}^{0}}$ distribution. The difference between the
efficiencies before and after mixing, 0.1\%, is considered to be the
systematic uncertainty. The total uncertainties associated with  the  efficiency for $\chi_{c0}$, $\chi_{c1}$, and $\chi_{c2}$ are 1.0\%, 0.9\%, and 0.4\%, respectively.

The systematic uncertainty due to the signal line shape is considered by changing the damping factor from
  $\exp(-E_\gamma^2/8\beta^2)$
  to $\frac{E_{0}^{2}}{E_{0}E_{\gamma}+(E_{0}-E_{\gamma})^2}$
 used by KEDR~\cite{KEDR}, where $E_{0}=\frac{m_{\psi(3686)}^{2}-m_{\chi_{cJ}^2}}{2m_{\psi(3686)}}$ is the peak energy of the transition photon, the differences in the fit results for $\chi_{c0}$, $\chi_{c1}$, and $\chi_{c2}$, 1.4\%,  1.9\%, and  0.4\% are assigned as the systematic uncertainties.

  The uncertainty associated with the detector resolution is studied by making the width of the Gaussion function to be free,  no changes are found for the $\chi_{c0}$, $\chi_{c1}$, and $\chi_{c2}$ signal yields, thus these uncertainties are neglected.

  The systematic uncertainties due to the $\chi_{c0}$, $\chi_{c1}$, and $\chi_{c2}$ mass and width in the fit are studied by changing them from free to the world average values~\cite{PDG2016}. The differences of the $\chi_{c0}$, $\chi_{c1}$, and $\chi_{c2}$ signal yields, 3.0\%, 0.4\%
and 3.9\%  are taken as the systematic uncertainties.

The uncertainty from the determination of $\chi_{cJ}$  signal events due to the fit range is obtained from
the maximum difference in the fit results by changing the fit range from  [3.30, 3.60] GeV/$c^{2}$ to  [3.30, 3.65] GeV/$c^{2}$ or [3.25, 3.60] GeV/$c^{2}$. The maximum differences in the fitted yields for $\chi_{c0}$, $\chi_{c1}$, and $\chi_{c2}$ are 0.9\%, 1.4\%, and 0.8\%, respectively.

The uncertainty due to the estimation of the background contribution using the $K_{S}^{0}$ and $\Sigma^{+}$ mass sidebands can be estimated by changing the sideband ranges. Changing the mass range of $ K^{0}_{S}$ from [0.466, 0.482], [0.514, 0.530] GeV/$c^{2}$ to [0.464, 0.480], [0.516, 0.532] GeV/$c^{2}$, and the mass range of $\Sigma^+$ from [1.1094, 1.1494], [1.2294, 1.2694] GeV/$c^{2}$ to [1.1074, 1.1474], [1.2314, 1.2714] GeV/$c^{2}$, and  varying  the non-$K_{S}^{0}$, non-$\Sigma^{+}$ mass region accordingly, the differences of $\chi_{c0}$, $\chi_{c1}$, and $\chi_{c2}$ signal yields are 0.3\%, 0.1\%, and 0.5\%, respectively. The uncertainty from the  shape of the non-$\chi_{cJ}$  background is estimated by changing the polynomial degree from the first to the second in fitting the $\Sigma^+\bar{p}K_{S}^{0}$ invariant mass,  and the  differences in the fit results are 2.8\%, 1.4\%, and 1.4\%, respectively. The total uncertainties associated with the background shape are  2.8\%, 1.4\%, and 1.5\% for $\chi_{c0}$, $\chi_{c1}$, and $\chi_{c2}$ decays, respectively.

The systematic uncertainties due to the secondary branching fractions of
$\psi(3686) \to \gamma\chi_{c0}~(\chi_{c1}, ~\chi_{c2}),~\Sigma^{+}\to p \pi^{0},~K_{S}^{0}\to \pi^{+}\pi^{-}$, and $\pi^{0}\to\gamma\gamma$ are 2.0\% (2.5\%, 2.1\%), 0.6\%, 0.07\% , and 0.03\%~\cite{PDG2016} respectively. Therefore, the uncertainties of the secondary branching fractions are 2.1\%, 2.6\% and 2.2\% for  $\chi_{c0}$, $\chi_{c1}$, and $\chi_{c2}$ decays, respectively.

The number of  $\psi(3686)$ events is determined to be
$(448.1\pm2.9)\times 10^6$~ by
 counting inclusive hadronic events from  $\psi(3686)$ decays~\cite{psi'data}, thus the uncertainty is about 0.6\%.

The total systematic uncertainty is the sum in quadrature of all uncertainties added for each $\chi_{cJ}$ decay.
  \begin {table*}[htp]
    \centering {\caption {Systematic uncertainty sources and their
        contributions (in \%). }
    \label{summary_of_syserr}}
    \begin{tabular}{l|c|c|c}  \hline \hline
       Source &  $\mathcal{B}(\chi_{c0} ) $& $\mathcal{B}(\chi_{c1} )$&$\mathcal{B}(\chi_{c2} )$\\ \hline
      Photon detection & 3.0 &3.0& 3.0\\ \hline
       PID and tracking & 0.6  &0.5&0.4\\ \hline
      4C  kinematic fit & 0.4  &0.4&0.3\\ \hline
      $K_{S}^{0}$ reconstruction &1.2&1.2&1.2\\ \hline
      $\pi^{0}$  mass window & 0.3  & 0.3& 0.3\\ \hline
      $K_{S}^{0}$  mass window & 0.3  & 0.3& 0.3\\ \hline
      $\Sigma^{+}$  mass window & 0.1  & 0.1& 0.1\\ \hline
       Efficiency  &  1.0  &  0.9&  0.4\\ \hline

      Signal line shape & 1.4  & 1.9 & 0.4\\ \hline
      Mass and width of $\chi_{cJ}$ & 3.0  & 0.4 & 3.9\\ \hline
       Fit range & 0.9  & 1.4& 0.8\\ \hline
      Background shape & 2.8  & 1.4& 1.5\\ \hline
      Intermediate decay &2.1  &2.6&2.2 \\ \hline
      Number of $\psi(3686)$ & 0.6  &0.6&0.6\\ \hline
      Total & 6.0 &  5.2& 5.9\\ \hline\hline
    \end{tabular}
\end{table*}

\section{Summary}
Using the $(448.1 \pm 2.9)\times$10$^{6}$ $\psi(3686)$ events
accumulated with the BESIII detector, the study of $\chi_{cJ}\to
\Sigma^{+}\bar{p}K_{S}^{0}~(J = 0, 1, 2)$ is performed for the first
time, and clear $\chi_{cJ}$ signals are observed.  The branching
fractions of $\chi_{cJ}\to \Sigma^{+}\bar{p}K_{S}^{0}$ are determined
to be $(3.52 \pm 0.19\pm 0.21)\times10^{-4}$, $(1.53\pm 0.10\pm
0.08)\times10^{-4}$, and $(8.25 \pm 0.83 \pm 0.49)\times10^{-5}$ for
$J=0$, 1, and 2, respectively, where the first and second uncertainty
are the statistical and systematic ones, respectively.  Due to the
limited statistics, no evident structure is observed in the invariant
mass of any subsystem.

\begin{acknowledgments}
The BESIII collaboration thanks the staff of BEPCII and the IHEP computing center for their strong support. This work is supported in part by National Key Basic Research Program of China under Contract No. 2015CB856700; National Natural Science Foundation of China (NSFC) under Contracts Nos. 11625523, 11635010, 11735014, 11822506, 11835012; the Chinese Academy of Sciences (CAS) Large-Scale Scientific Facility Program; Joint Large-Scale Scientific Facility Funds of the NSFC and CAS under Contracts Nos. U1532257, U1532258, U1732263, U1832207; CAS Key Research Program of Frontier Sciences under Contracts Nos. QYZDJ-SSW-SLH003, QYZDJ-SSW-SLH040; 100 Talents Program of CAS; INPAC and Shanghai Key Laboratory for Particle Physics and Cosmology; ERC under Contract No. 758462; German Research Foundation DFG under Contracts Nos. Collaborative Research Center CRC 1044, FOR 2359; Istituto Nazionale di Fisica Nucleare, Italy; Ministry of Development of Turkey under Contract No. DPT2006K-120470; National Science and Technology fund; STFC (United Kingdom); The Knut and Alice Wallenberg Foundation (Sweden) under Contract No. 2016.0157; The Royal Society, UK under Contracts Nos. DH140054, DH160214; The Swedish Research Council; U. S. Department of Energy under Contracts Nos. DE-FG02-05ER41374, DE-SC-0010118, DE-SC-0012069; University of Groningen (RuG) and the Helmholtzzentrum fuer Schwerionenforschung GmbH (GSI), Darmstadt.
\end{acknowledgments}


\end{document}